\documentstyle[epsf,prd,aps,epsf]{revtex}
\begin{document}

\newcommand{\gsim}{\gtrsim}
\newcommand{\lsim}{\lesssim}
\newcommand{\psim}{\mbox{\raisebox{-1.0ex}{$~\stackrel{\textstyle \propto}
{\textstyle \sim}~$ }}}
\newcommand{\vect}[1]{\mbox{\boldmath${#1}$}}
\newcommand{\lmk}{\left(}
\newcommand{\rmk}{\right)}
\newcommand{\lnk}{\left\{ }
\newcommand{\nn}{\nonumber}
\newcommand{\rnk}{\right\} }
\newcommand{\lkk}{\left[}
\newcommand{\rkk}{\right]}
\newcommand{\lla}{\left\langle}
\newcommand{\p}{\partial}
\newcommand{\rra}{\right\rangle}
\newcommand{\beq}{\begin{equation}}
\newcommand{\eeq}{\end{equation}}
\newcommand{\beqa}{\begin{eqnarray}}
\newcommand{\eeqa}{\end{eqnarray}}
\newcommand{\lab}{\label}
\newcommand{\sol}{M_\odot}
\newcommand{\mch}{{\cal M}}
\newcommand{\vex}{{\vect x}}
\newcommand{\vel}{{\vect \Omega}_l}
\newcommand{\ven}{{\vect \Omega}_s}
\newcommand{\vep}{{\vect p}}
\newcommand{\veq}{{\vect q}}
\newcommand{\veo}{{\vect \Omega}}

%\if0
\draft
\title{Effects of finite arm-length of LISA on analysis of  gravitational waves from
MBH binaries}
\author{Naoki Seto
%\footnote{e-mail: seto@vega.ess.sci.osaka-u.ac.jp}
}
\address{Department of Earth and Space Science, Osaka
University, Toyonaka 560-0043, Japan}

\maketitle

\begin{abstract}
 Response of an interferometer becomes complicated for gravitational wave 
 shorter than the  arm-length of the detector, as nature of wave
 appears strongly. We have studied how parameter estimation  for
 merging massive black hole binaries are affected by this complicated
 effect in the case of LISA. It is shown that three dimensional positions
  of some binaries might be determined much  better than the past
 estimations that use the long wave approximation. For equal mass binaries
 this improvement is most prominent at $\sim 10^5\sol$.
\end{abstract}

%\pacs{PACS number(s): 95.55.Ym 04.80.Nn, 98.62.-g }
%\fi

%%%%%%%%%%%%%%%%%
\section{Introduction}
%%%%%%%%%%%%%%%%5

The {\it Laser Interferometer Space Antenna} (LISA), a joint project of NASA
and ESA,  is planed to be launched  in 2011 \cite{lisahttp}.  LISA
is sensitive to low frequency gravitational waves at 
$10^{-4}{\rm Hz}\lsim f \lsim  10^{-1}$Hz,  and would directly  confirm 
gravitational radiation from some known Galactic binaries
\cite{lisa}. Thousands of closed white dwarf binaries would be also
detected by LISA.
They can be 
regarded as insurances of the project, but other exciting phenomena might
be observed \cite{lisa}.  
Coalescence of a massive black hole (MBH)  binary is  the most
spectacular and energetic event. 
 Gravitational wave from a MBH binary is  in  itself
very interesting for studies of  general relativity, but would also
bring significant 
impacts on astrophysics and cosmology \cite{lisa}, even though  event
rate is highly 
unknown at present \cite{Haehnelt:wt}. Parameter estimation
errors are 
basic measures for   discussion of gravitational wave
astronomy. For example, with small error box for the position  of a MBH
binary we might specify its host galaxy and investigate it with various
observational tools. 

LISA is constituted by three space crafts that keep a triangle
configuration with side (arm-length) $l=5.0\times 10^6$km and trail
$\sim 20^\circ$ behind the Earth \cite{lisa}. 
Response of a detector  becomes complicated for   gravitational waves with
wave-length  $\lambda\lsim l$, as the nature of wave appears
strongly \cite{Larson:1999we,Schilling:id,Cornish:2002rt}. 
This interesting feature might 
be an advantage for signal analysis but has not been taken in properly
so far. In 
this paper we discuss how the parameter estimation errors are changed by
this  effect.

This paper is organized as follows. In Section II we study the effect of
a finite arm-length on the phase shift $\delta \phi$ of the detector,
and introduce 
the past analysis that is based on the simple picture using the
variations of the arm-lengths $\delta l$ under the long wave
approximation. In Sec III A gravitational wave forms  from MBH binaries are
briefly discussed. Noise spectrum  for both the variations $\delta l$
and the phase sifts $\delta \phi$ are presented in Sec III B. The
spectrum for the latter is simply reproduced from that for the former.
 In Sec IV our numerical results are shown, and signal to noise ratio
and the parameter estimation errors are compared for the two methods (with
and without the long wave approximation). Though we mainly study
gravitational waves from MBHs,  results for nearly
monochromatic  sources (from {\it e.g.} Galactic binaries) are also presented.
Sec V is devoted to  discussion. In Appendix A we make same kind of 
comparisons for the noise canceling combination of data streams.
 In Appendix B we present explicit expressions for detector's response
 to
gravitational wave from a binary.

%%%%%%%%%%%%%%%%%
\section{Detectors with finite arm-length}
%%%%%%%%%%%%%%%%5

We  name  the three vertexes (space crafts) of LISA as A, B and C.
Cutler \cite{Cutler:1998ta} studied   parameter estimation of binaries
from variations of arm-lengths in the following forms
\beq
\delta l_{AB}(t)-\delta l_{AC}(t),~~~
\lnk(\delta l_{BA}(t)-\delta l_{BC}(t))-(\delta l_{CB}(t)-\delta
l_{CA}(t))\rnk/\sqrt{3}.\label{ldata} 
\eeq
The normalization factor $1/\sqrt{3}$ in the second expression is
explained later. 
Here $\delta l_{XY}$ is the variation of the arm-length between two space
crafts $X$ and $Y$ ($X,Y\in\{A,B,C\},~X\ne Y$). This simple picture for detector's
response  is valid 
only at long wave  limit  where gravitational wave-length is much larger
than the detector's arm-length (see also
\cite{Hughes:2001ya,Moore:1999zw}). 
 With quantities more close to
detector's output  \cite{lisa} these 
two data should be replaced by 
\beq
\delta\phi_{AB}(t)-\delta\phi_{AC}(t),~~~
\lnk(\delta\phi_{BA}(t)-\delta\phi_{BC}(t))-(\delta\phi_{CB}(t)-\delta\phi_{CA}(t))\rnk/\sqrt{3},\label{data}
\eeq
where $\delta\phi_{XY}$ represents the single-arm phase shift of the laser beam
 that leaves  a detector (vertex of the triangle) X for Y and then
 returns back to 
 X.    The
 quantity $\delta\phi_{AB}-\delta\phi_{AC}$  is an interferometer
 signal obtained at the detector A.

 In Ref.\cite{Cutler:1998ta} it is assumed
 that the
 noise for the time variations $\delta l_{XY}$ is 
 stationary, Gaussian and symmetric among three arms. 
We make same assumption for the corresponding  phase shifts  $\delta \phi_{XY}$
In this case 
 noises  of the above two data  (\ref{data}) (or (\ref{ldata})) do not
 correlate to each other. 
%As in past studies for parameter estimation of binaries
%\cite{Cutler:1998ta,Hughes:2001ya,Moore:1999zw}, we do not use the noise 
%canceling combinations of data (see {\it e.g.} \cite{Armstrong:uh}).

Response of the phase shift  $\delta\phi_{AB}$ for 
gravitational wave $h$  incoming from 
a direction $\veo_s$ is expressed as follows
\cite{Larson:1999we,Schilling:id,Hellings:jm} 
\beq 
\frac1{2\pi \nu_0}\frac{d\delta\phi_{AB}(t)}{dt}=\frac12 \cos2\psi_{AB}
[(1-\cos\theta_{AB})h(t,A)+2\cos\theta_{AB}
h(t-\tau,B)-(1+\cos\theta_{AB})h(t-2\tau,A)],\label{pt}
\eeq
where  $\theta_{AB}$ is the
angle between the source direction $\ven$ and the arm 
$\vex_A-\vex_B$, and $\nu_0$  is the fundamental frequency of the
laser.  $\psi_{AB}$ is the principle polarization angle of the quadrupole
gravitational wave $h$ for the arm $\vex_A-\vex_B$. We denote the
polarization basis tensor $H_{ab}$ 
of the wave as $H_{ab}=p_ap_b-q_aq_b$ using two orthogonal  vectors $\vep$
and $\veq$ with $\vep\cdot\ven=\veq\cdot\ven=\vep\cdot\veq=0$ and $|\vep|=|\veq|=1$. Then the angle $\psi_{AB}$ 
is given by $\tan \psi_{AB}=({\veq}\cdot(\vex_A-\vex_B))/({\vep}\cdot(\vex_A-\vex_B))$ \cite{Hellings:jm}.  
The quantity 
$h(t,A)$ is value of the  gravitational wave at point A and
time $t$,  and  can be
expressed  as $h(t+\veo_s\cdot\vex_A(t)/c)$ in the present case. 
We have denoted  the  propagation time of light for the
 arm $l$ by  $\tau\equiv l/c$, 
  and neglected very small relative motions between the vertexes in the
time scale $\tau$.  In Appendix B we present explicit expression of
Eq.(\ref{pt}) for a binary source.

When the gravitational wave-length    is much larger
than the arm-length $l$, namely $f\ll (2\pi\tau)^{-1}\equiv f_{arm}$, 
 structure of wave is irrelevant at the spatial scale $l$ and the wave $h$
 causes variations $\delta \phi_{AB}$ or $\delta l_{AB}$ in very simple
 manners. In this long wave limit
the phase
shift is directly proportional to 
the variation of the arm-length $\delta l_{XY}$ as $\delta \phi_{XY}/2\pi\nu_0=
\delta l_{XY}/2c$,  and   two data sets (\ref{ldata}) and
(\ref{data}) are equivalent.    We
have the 
following  relation by perturbative evaluation of Eq.(\ref{pt})
(see also Ref.\cite{Hellings:jm}) 
\beq
\frac{\delta \phi_{AB}}{2\pi \nu_0\tau}=
\sin^2\theta_{AB} \cos 2\psi h(t)=\frac{\delta l_{AB}}{2l}\label{lw}. 
\eeq
LISA has arm-length  $l=5.0\times 10^6$ km,  corresponding to the
critical
frequency
 $f_{arm}=0.01$Hz.
Note that the frequency $f$  does not appear in the
above relation 
(\ref{lw}).    
In this long wave limit  the angular pattern function
(usually 
denoted as $F^{+,\times}$ \cite{thorne}) contains complete information of the
angular dependence 
of the interferometer signal ({\it e.g.} 
$\delta\phi_{AB}-\delta\phi_{AC}$).  Two data
streams (\ref{ldata}) are equivalent to  responses of two
$90^{\circ}$-interferometers 
rotated by $45^\circ$ (see discussion in  \cite{Cutler:1998ta}). The factor $1/\sqrt{3}$
in the second expression of (\ref{ldata}) or (\ref{data}) is given for
this purpose.

When the gravitational wavelength $c/f$ is comparable to or shorter
than the arm-length $l$, the phase shifts $\delta\phi_{XY}$ show
complicated response that 
depends strongly on the frequency $f$ and the source direction relative to
the detector (see Appendix B). As discussed earlier,  this troublesome but interesting
 effects have been cut down so far in analysis of parameter
estimation errors for binary  sources.
 Past analysis was
performed in the following manner
\cite{Cutler:1998ta,Hughes:2001ya,Moore:1999zw}. 

(i) The angular averaged transfer
function $T(f)$ between  the phase shift $\delta \phi$  and wave amplitude
$h$ was used $\delta \phi\propto T(f)h$ to include amplitude modulation
caused by finiteness of the arm-length $l\ne 0$.
 Its effect can be  effectively
absorbed in the noise curve $\sqrt{S_h(f)}$ for the wave amplitude
$h$. 

(ii) Then the simple analysis 
with the variation of the arm-length $\delta l_{XY}$ (valid only at the
long wave limit) was  
applied  with the angular pattern function $F^{+,\times}$ for 
the angular dependence of the 
response (and also Doppler phase caused by velocity of detectors).

At
high frequency limit  $f\gg f_{arm}$ the 
transfer function becomes   $T(f) \propto f^{-1}$ due to cancellation
of wave within  arms. The measurement sensitivity of  LISA for the
phase shift 
$\delta \phi$  is $\sqrt{S_{\delta\phi}(f)}\propto f^0$ (in
units of  ${\rm Hz}^{-1/2}$) at $10^{-2}{\rm Hz}\lsim f\lsim 10^0{\rm Hz}$  where the shot
noise is dominant source of noise. Thus LISA has
$\sqrt{S_{h}(f)}\propto \sqrt{S_{\delta\phi}(f)}T(f)^{-1}\propto f$
 at high
frequency region \cite{lisa,Larson:1999we,Schilling:id}.

Here we discuss basic aspects of signal analysis with paying attention
to the finiteness of the arm-length $l$.
The most important quantity for detection of gravitational wave is the signal
to noise ratio (SNR) \cite{thorne}. For its estimation the above mentioned method with
the 
angular averaged transfer function  would be effective for LISA
considering its rotation and frequency average of chirping signals.  But
parameter estimation errors ({\it 
e.g.} error box for direction of a source) might become much smaller
than the previous 
analysis when we include the complicated response of the phase shift
as it is.
Gravitational  wave signal measured  by a detector (as  expression
(\ref{data}))  is given by  values of  gravitational wave $h$ at 
different positions and times. Time delay
between them would  also  affect parameter estimation.
These two  effects, time delay and cancellation within the arms,  are
natural outcome of 
the finiteness of the  arm-length and appear in a similar form, namely, linear
combination of terms like $\exp[2\pi i f \veo_s\cdot
(\vex_A-\vex_B)/c]$ in Fourier space of the  time variable $t$.
Our primary aim in this paper is to show how the parameter estimation would
be changed if we  simply
replace the arm-length variation   $\delta l_{XY}$ (as in Eq.(\ref{ldata})) with the corresponding
phase shifts  $\delta \phi_{XY}$  (as in Eq.(\ref{data})) without using the long wave
approximation. 

\section{Analysis of Gravitational Waves from Merging Massive Black Hole
 Binaries}
\subsection{Gravitational Waveform}

We investigate the parameter estimation errors expected in matched
 filtering method \cite{Finn:1992wt}. 
When the wave signal contains fitting parameters
$\lnk \gamma_i\rnk$, their  estimation errors
 (variances)  for signal analysis are evaluated 
with using  the Fisher's
information matrix  $\Gamma_{ij}$ as $\lla \Delta\gamma_i \Delta\gamma_j
\rra=\Gamma^{-1}_{ij}$. 
 We analyze the
phase shifts (\ref{data})  by extending earlier studies 
to include effects caused by the arm-length, but other points are almost
 identical to Ref.\cite{Cutler:1998ta} that uses the variations
 (\ref{ldata}) under the long wave approximation. 
%The inner product $(a|b)$ of two data $a(t)$ and $b(t)$ is
%defined as $(a|b)=4{\rm Re} \int_0^\infty df \frac{{a^*(f)b(f)}}{S(f)}$ 
Detailed analysis for the phase shift is given in Appendix B.

%For multiple signal $h_m$ with uncorrelated noise the
%above inner product $(h|h)$ are  replaced by  summation $\sum_m
%(h_m|h_m)$.

For  gravitational wave emitted by  a binary we adopt the
restricted post-Newtonian approach, but higher-order harmonics could
become important in some cases \cite{Hellings:2002si}.   We use the 1.5PN phase for a
circular orbit  \cite{Cutler:1993tc}
\beq
\Psi(f)=2\pi f t_c-\phi_c-\frac\pi4+\frac34(8\pi G c^{-3} \mch_z f)^{-5/3} \lkk
1+\frac{20}9 \lmk \frac{943}{336}+\frac{11\mu_z}{4M_z}\rmk
x+(4\beta-16\pi) x^{3/2}  \rkk, \label{pev}
\eeq
where $t_c$ and $\phi_c$ are integration constants, $\beta$ is the
spin-orbit coupling term, $\mu_z$, $M_z$ and $\mch_z$ are the reduced
mass, the total mass and the chirp mass of the binary. All of the mass
parameters are multiplied by the factor $(1+z)$ ($z$: redshift of  the
binary) with suffix $z$. The total mass $M_z$ is expressed by other two
masses as $M_z=
\mch_z^{5/2} \mu_z^{-3/2}$.
The post-Newtonian expansion parameter  $x$ is defined as $x\equiv \lnk
G \pi c^{-3}
M_z f 
\rnk^{2/3} $.
We have used the stationary phase approximation for the Fourier
transformed waveform.   
In our analysis total number of the fitting parameters is 10 as
$\mch_z,\mu_z,\beta,\phi_c, t_c, \ln D$(luminosity distance) and 
$(\theta_s,\phi_s),(\theta_l,\phi_l)$. The latter four parameters define
the direction $\ven$ and the orientation $\vel$ of the binary in
a fixed 
frame at  the barycentre of the solar system. In this paper we use the
error ellipse 
\beq
\Delta \veo_i\equiv 2\pi \sin \theta_i\sqrt{\lla
\Delta\theta_i^2\rra \lla \Delta\phi_i^2\rra -\lla
\Delta\theta_i\Delta\phi_i\rra^2 },\label{angl}
\eeq
 for the angular 
parameters ($i=s,l$) \cite{Cutler:1998ta} and $\Delta V\equiv \Delta D \Delta \ven$ for
the three dimensional position error of a source. We also fix $\beta=0$ for
the 
true value of the parameter $\beta$.

Unless stated explicitely, we put the upper cut-off frequency $f_{cut}$
of signal integration at  $f_{isco}$ 
 when
the binary separation becomes 
$r=6M_z/c^2(1+z)$,  roughly corresponding to the inner most stable circular
orbit.
The  observed wave frequency at this separation becomes 
\beq
f_{isco}=\frac{6^{-3/2}c^3}{G\pi M_z}=0.022\lmk \frac{M_z}{2\times10^5
M_\odot}\rmk^{-1} 
{\rm Hz}\label{fcut}.
\eeq
We start integration of the wave signal  from the time when 
the binary is 1yr
before its coalescence in the observer's frame.
The gravitational wave  
frequency $f_0$ at this starting time is given as follows
\beq
f_0=1.9\times10^{-4} \lmk\frac{\mch_z}{0.87\times 10^5 M_\odot} \rmk^{-5/8}{\rm Hz}. \label{f0}
\eeq

\subsection{Noise Curve}

We make quantitative analysis using noise curve of LISA that is sum of
the instrumental noise and the binary confusion noise. For the
instrumental noise
spectrum  given for the wave amplitude $h$ (almost equivalently for
$\delta l$) we 
adopt the 
following function \cite{Cutler:1998ta}
\beq
S_i(f)=5.049\times 10^5[\alpha_1(f)^2+\alpha_2(f)^2+\alpha_3(f)^2]~{\rm
Hz}^{-1}\label{sp}, 
\eeq 
where $\alpha_1(f)=10^{-22.79}(f/10^{-3}{\rm Hz})^{-7/3}$ is mainly the acceleration
noise, $\alpha_2(f)=10^{-23.04}$ is mainly the shot noise,  and
$\alpha_3(f)=10^{-24.54}(f/10^{-3}{\rm Hz})$ approximately 
represents the cancellation effects
due to the finiteness of the  arm-length. The last term reflects the angular
averaged transfer function as explained before, and becomes important at
$f\gsim 
10^{-2}$Hz (see {\it e.g.} \cite{Larson:1999we}). We also include the
binary confusion noise $S_c(f)$ (estimated for 
one year observation) that is stronger than the instrumental one at
$f\lsim 10^{-2.5}$Hz and shows strong dependence on  frequency
$f$ \cite{Bender:hs}. Around $f\gsim 10^{-2.75}$Hz the confusion noise
 decreases
significantly as the Galactic 
binaries are resolved at higher frequencies.
Its explicite expression is given as follows \cite{Cutler:1998ta}
\beq
S_c(f)=\cases{
10^{-42.685}f^{-1.9}{\rm Hz^{-1}}, ~~ f\le 10^{-3.15},  \cr
10^{-60.325}f^{-7.5}{\rm Hz^{-1}}, ~~ 10^{-3.15}\le f\le 10^{-2.75}, \cr
10^{-46.85}f^{-2.6}{\rm Hz^{-1}}, ~~   10^{-2.75}\le f, \cr
}
\eeq
where $f$ is written in units of Hz.

Now let us simply recover the noise spectrum $S_{\delta \phi}(f)$ for
the phase shift $\delta 
\phi_{AB}-\delta
\phi_{AC}$ from that for the variation $\delta l$. 
%The above spectra ({\it
%e.g.} Eq.(\ref{sp}))
%are given for the gravitational wave amplitude $h$ or $(\delta
%l_{AB}-\delta l_{AC})/l$, but can be easily 
%related to the spectra  $S_{\Delta\phi}(f)$ of the   phase shift by noticing
%their behaviors 
At low frequency region $f\ll f_{arm}
$ they are simply related to each other as  expressions (\ref{ldata})
(\ref{data}) and are essentially equivalent (see Eq.(\ref{lw})). We use the
following functional 
shape for  the instrumental noise of 
the phase shift $S_{\delta\phi}(f)$ as
$ S_{\delta\phi}(f)\propto (\alpha_1(f)^2+\alpha_2(f)^2)$ by removing
the effect of the angular averaged transfer function from
$S_i(f)$. Note 
that the confusion noise is
expected to be much smaller than the instrumental one at the relevant
frequencies $f\gsim f_{arm}$. 
For the second data of expression  (\ref{data}) we use the same noise
curve as the first one. 

%Strictly speaking this is not correct but would
%not change general behaviors of our results.

%The two data streams (\ref{data})  are integrated for the 
%last 1 year before 
%coalescence upto the cut off frequency given by Eq.(\ref{fcut}). 
Our
calculation 
using the phase shift is much more complicated than the past one.
 But our numerical results should  coincide with these
 by  past one when we 
decrease  the arm-length $l$ and use the shape of the noise curve
$S_h(f)$ instead of $S_{\delta\phi}(f)$. We have confirmed that the
parameter errors $\Delta \gamma_i$ given in table 2 of
Ref.\cite{Cutler:1998ta} (obtained 
by past simple calculation) is reproduced with this procedure.

%%%%%%%%%%%%%%%%%
\section{results}
%%%%%%%%%%%%%%%%5
We have analyzed gravitational waves from MBH binaries with 300 
realizations of random  directions $\ven$ and orientations $\vel$.
Firstly, their SNRs and estimation errors of the fitting parameters $\lnk
\gamma_i \rnk$ are 
calculated in both (i) past approach (suffix $0$) with angular
averaged transfer function under the long wave approximation, and (ii)
our method (suffix $L$) based on the 
phase shifts of expression (\ref{data}). In figure 1 we present
distribution for 
ratios of SNRs by two methods  $(SNR)_L/(SNR)_0$ and  position errors
$\Delta V_L/\Delta V_0$, $(\Delta\ven)_L/(\Delta\ven)_0$ and
$\Delta D_L/\Delta D_0$ obtained  
for equal mass MBH binaries with redshifted masses 
$M_{z}=10^4+10^4, 10^5+10^5$ and $10^6+10^6\sol$.
Note that these ratios do not depend on cosmological parameters or
distance $D$ to the MBH binaries. As shown in the bottom panel of figure
1, ratios 
of SNRs are close to unity and difference of two methods  is very small
(within $20\%$).
However, two methods show considerable difference for the position errors.  For $10^5+10^5\sol$ MBH binaries the errors $\Delta V_L $
become typically 10 times smaller than the past  estimate $\Delta
V_0$. For some samples the errors $\Delta V$ are reduced even by a
factor $\sim 10^{-2}$. Comparing errors $\Delta \ven$ and $\Delta D$,
 the former become smaller by the finiteness of the arm-length than
the latter.  But note that the angular resolution $\Delta \ven$ is,
roughly speaking, a product of two  errors $\Delta \theta_s$ and $\Delta
\phi_s$ as in Eq.(\ref{angl}).

Interestingly enough, difference between two methods is not a monotonic
function 
with respect to BH mass and is most prominent around $\sim
10^5\sol$. This mass dependence can be understood as follows. At higher
masses $\gg 10^5\sol$ the upper cut-off frequency $f_{isco}$ is smaller than the critical
frequency $f_{arm}$. Therefore nothing is different between two methods
with expressions (\ref{ldata}) and (\ref{data}). With the quadrupole formula
 for gravitational radiation
 the time
before coalescence is given by the frequency $f$ and the chirp mass $\mch_z$
as follows \cite{thorne}
\beq
t_{GW}=8.4\times 10^4 \lmk \frac{f}{10^{-2.75}{\rm Hz}}\rmk^{-8/3} \lmk
\frac{\mch_z}{0.87\times 10^5 \sol}\rmk^{-5/3} {\rm sec}.
\eeq
For lower mass BH binaries ($\ll 10^5\sol$) LISA moves longer than the
arm-length $l$ during 
the phase $f\gsim 10^{-2.75}$Hz where signal becomes very strong due to
decrease of the binary 
confusion noise. Thus effective baseline of the detector becomes larger
than the arm-length $l=5.0\times 10^6$km for smaller mass BH binaries
and impact of the finite arm-length would decrease.

Now we  investigate various aspects of parameter estimation caused by the
finiteness of the arm-length using a specific set of binary parameters.
We pick up the binary that has the smallest volume ratio $\Delta
V_L/\Delta V_0=0.011$ in 300 realization of figure 1 for $10^5+10^5\sol$
MBH binaries.
It has angular parameters $\theta_s=2.49, \phi_s=0.03, \theta_l=2.32$
and $\phi_l=4.46$.  Here we present the estimation errors $\Delta \gamma_i$,
not ratios as in figure 1. To normalize the amplitude of the signal we
take the redshift of MBH binaries at $z=1$ with cosmological parameters
$\Omega_0=0.3, \lambda_0=0.7$ and $H_0=75$km/sec/Mpc. In figure 2 estimation
errors $\Delta \ven, \Delta \vel, \Delta D/D$ and $\Delta
\mu_z/\mu_z$ are presented as  functions of the upper cut-off frequency
$f_{cut}$ that was fixed at $f_{isco}$ with Eq.(\ref{fcut}) in the case
of figure 1.  We fix the lower cut-off frequency at $f_0$ (eq.[\ref{f0}])

We have found that the intrinsic binary parameters such as 
$\mch_z, \mu_z, \beta, t_c$ and $\phi_c$ depend weakly on the cut-off frequency
$f_{cut}$.  
However, errors for the binary position $\Delta\ven$, $\Delta D/D$
or  its orientation $\Delta \vel$ decrease significantly at
$f_{cut}\gsim 0.01$Hz where the long wave approximation breaks down. This shows remarkable contrast to the  past analysis
shown with thin lines.
 Our results seem reasonable, as the response of a detector with
finite arm-length depends strongly on the direction of the source
$\ven$, and information  
of the distance $D$  or orientation $\vel$ is tightly correlated to them
(Appendix B).
Significant reduction of position errors $\Delta V_L$ at higher
frequencies would give further motivation for studies of nonlinear
gravitational dynamics ({\it e.g.} Post-Newtonian approach)
that is a very tough problem on general relativity.  By analyzing
gravitational wave close to the 
final coalescence we might identify the host galaxy of a MBH
binary!

Next let us make hypothetical experiments to clarify some interesting
points. We use the same set of the parameters as
in figure 
2 with $f_{cut}$ given by Eq.(\ref{fcut}). As commented before, the finiteness of the arm-length causes two
similar effects (i) cancellation of waves within the arm and (ii) time
delay between the vertexes of LISA. To extract effects only of the latter we
calculate the volume error $\Delta V_{L'}$ using data from three
interferometers that exist at three vertexes of LISA but have arm-length
$l\to0$ with angular averaged transfer function for sensitivity of
$h$. Thus only the 
positions (separation) of the detectors are different from the past
analysis  that take the separation $l=0$. We
obtain $\Delta V_{L'}/\Delta V_0=0.15$. This result indicates that two
effects work cooperatively.
Next we stop motion of LISA and keep its position at time $t=t_c$. In
this case the volume error  $\Delta V_{L''}$ becomes $\Delta
V_{L''}/\Delta V_{0}=0.012$ and is very close to $\Delta
V_{L}/\Delta V_{0}=0.011$ that includes motion of LISA. In the past
analysis (with $l=0$) we use  the
amplitude modulation through the pattern function and the Doppler phase
modulation both caused by motion of LISA \cite{lisa},  and  cannot solve degeneracy
of sources direction  $\ven$ and 
other variables when LISA stops. The response of a detector with $l\ne
0$ depends strongly both on angular variables and frequency of
incoming waves at $f\gsim f_{arm}$ (see Appendix B). Thus we can solve the degeneracy
for chirping binaries 
even without motion of LISA, though there would be two solution for
detector's signal due to the symmetry of source-detector configuration.
 This is a qualitatively interesting point.
%and might be also applied for a  ground-based detector  to
%determine the direction of a binary source.

Considering the mass dependence of our results, it is expected that
 the past simple method using expression
(\ref{ldata}) under the long wave approximation would be  effective  
for studying galactic compact binaries.  These binaries with $f\lsim
 0.1$Hz are nearly 
 monochromatic and 
have 
long durations $t_{GW}\gg 1$yr in the LISA band  due to their small chirp masses \cite{lisa}. 
We have investigated binaries with $m_1=m_2=1\sol$,  and the time to
 coalescence  $t_{GW}=30$yr and
 100yr. The wave frequency becomes $f=0.071$Hz at $t_{GW}=30$yr and
 $f=0.045$Hz at  $t_{GW}=100$yr.  We fix the observation period at 1yr,
 thus do not observe  the final
 coalescence in contrast to  the analysis for MBH binaries. We calculate
 parameter estimation errors in this situation with random direction $\ven$ and
orientation  $\vel$. It is found that the 
 differences between ($\Delta V_L$,$\Delta V_0$)  or ($(SNR)_L$,$(SNR)_0$) are  less
 than 15 percent. Recalling that we have made a simple treatment for the
 effects of 
 the transfer function, this result seems excellent.

%%%%%%%%%%%%%%%%%
\section{discussion}
%%%%%%%%%%%%%%%%5

In this paper we have studied how data analysis of gravitational waves
from MBH binaries are affected by finiteness of the arm-length of LISA
with using the phase shift of detectors. 
We have numerically confirmed that
the 
past method with the long wave approximation is very effective
for estimation of SNRs that are the most important quantities for
detection of gravitational waves.
However,  LISA is able to observe merging MBH binaries with significant SNRs
($\gsim 1000$) even at cosmological distances \cite{lisa,Cutler:1998ta,Hughes:2001ya,Moore:1999zw},  and aspects of
gravitational wave astronomy are more relevant for them, rather than
SNRs.  We have examined the parameter estimation errors $\Delta
\gamma_i$ expected in both 
two  methods, and shown that three dimensional position of MBH binaries
could be determined much better than the past estimations. In the case of
equal  mass binaries differences between two methods are most prominent
at $\sim 10^5\sol$ and the volume of the error box can  decrease
significantly. We have shown that (i)
this  reduction is mainly
cause by gravitational waves close to the final coalescence and (ii) position
of chirping binaries can be in principle, estimated even without motion of LISA.
The former would give further meaning for studies on strong
gravitational dynamics, and the latter makes remarkable contrast to
former discussions.  

\if0
Past analysis with using noise curve with angular averaged transfer
function might be more or less affected by inquiry for  detectability
(as represented by SNR) of the wave signal.  Gravitational wave
astronomy would  ripe in this century. We hope that this paper provides an
opportunity to take a fresh look at theoretical study of data analysis
for gravitational waves.
\fi

The author would like to thank an anonymous referee for helpful
comments to improve the manuscript.
This work was supported in part by
Grant-in-Aid of Scientific Research of the Ministry of Education,
Culture, Sports, Science and Technology  No. 0001416.

\appendix

\section{Noise Canceling Combination}
In the main text %we have studied how the parameter estimation changed
we have basically followed the model for signal  analysis in
Ref.\cite{Cutler:1998ta}, 
 and simply replace the variations (\ref{ldata}) with the corresponding
 phase shifts
 (\ref{data}) without resorting to  the long wave approximation.
It has been discussed recently that the laser phase noise can be removed
well by devising combination of data at different times
\cite{Armstrong:uh}.  This noise is 
caused by un-equal arm-lengths of space detectors in contrast to the
ground-based ones. 
In this appendix we discuss parameter estimation using this combination.
Let us consider the situation that the space crafts B
and C  coherently transmit laser beams back to the space craft A.
Then the following data stream $X(t)$ is a  noise canceling combination 
\beq 
X(t)=\delta\phi_{AB}(t)-\delta\phi_{AC}(t)-\delta\phi_{AB}(t-2l_{AC}/c)+\delta\phi_{AC}(t-2l_{AB}/c).  \label{xt}
\eeq
% 
%In section 4 we also discuss signal analysis with the data stream
%$X(t)$.
%
For quantitative study we take the limit $l_{AB}=l_{AC}=l=5.0\times
10^6$km.  Then the Fourier transformation of $X(t)$ is related to that for 
$\delta\phi_{AB}(t)-\delta\phi_{AC}(t)$ as
\beq
X(f)=(1-\exp(2\pi i f l/c))(\delta\phi_{AB}(f)-\delta\phi_{AC}(f)).\lab{xf}
\eeq
In this case the noise curve for the gravitational wave amplitude with data
$X$ is identical to  that with $\delta\phi_{AB}-\delta\phi_{AC}$
\cite{Larson:1999we}, if we only include the acceleration and shot noises.

In figure 3 we present the two ratios $(SNR)_L/(SNR)_0$ and $\Delta
V_L/\Delta V_0$ for the noise canceling combination  $X$ given in
Eq.(\ref{xt}). Note that the 
factor $(1-\exp(2\pi i l f/c))$ in Eq.(\ref{xf}) contains none of our ten fitting
parameters. Thus results for the data $X(t)$ is identical to that for
the single 
data stream $\delta\phi_{AB}-\delta\phi_{AC}$. Figure 3 shows same
kind of mass dependence as figure 1. Difference between $\Delta
V_L$ and $\Delta V_0$ is most prominent at mass $\sim 10^5\sol$ again.
But the ratio $\Delta 
V_L/\Delta V_0$ is generally  close to unity and effects of the finite
arm-length are smaller.

{\bf Note added} After submitting this paper, there appears
\cite{Prince:2002hp} that discusses three data streams $A,E,T$  
whose noises do not correlate. Our study can be easily  extend to
these data and more realistic results would be obtained.

\section{Phase shift of a detector for gravitational wave from binaries}
In this appendix we give an explicit expression for  the phase shift in
the form (eq.[3])
\beq 
\frac1{2\pi \nu_0}\frac{d\delta\phi_{AB}(t)}{dt}. \label{b1}
\eeq
The shift $\delta\phi_{AB}$  is defined for the laser beam  that leaves
the detector $A$ for $B$, and then returns back to $A$. For source of
gravitational radiation $h$ we consider a chirping binary at luminosity
distance $D$,  direction $\ven$, and orientation $\vel$. We denote
$\vex_Y$ as  the
position vector of a detector $Y\in \{A,B,C\}$,
and  use a coordinate system fixed to the barycentre of the solar
system. As we do not directly use the angular pattern functions $F_+$ and
$F_\times$ (see \cite{thorne}), there is no need to  introduce a
coordinate system fixed to detectors.

The inclination angle $i$ of the binary is given as 
\beq
\cos i=\ven\cdot\vel.\label{b2}
\eeq
We denote the propagation time of light for the arm-length
by $\tau\equiv|\vex_A-\vex_B|/c$, and define the angle $\theta_{AB}$
between the direction of a binary $\ven$ and the arm $\vex_A-\vex_B$ as
\beq
\cos \theta_{AB} =\frac{\ven\cdot (\vex_A-\vex_B)}{c\tau}.
\eeq

It is convenient to use the principle axes $(\vep,\veq)$ for analysis of
gravitational wave from a binary. These two vectors are expressed in
terms of direction $\ven$ and orientation $\vel$ as
\beq
\vep=\frac{\ven\times\vel}{|\ven\times\vel|},~~~\veq=-\ven\times \vep. \label{b4}
\eeq
With these vectors gravitational wave from a binary is decomposed to two
polarization ($+$ and $\times$) modes whose phases differs
by $\pi/2$. 
The plus ($+$) mode has polarization basis tensor 
$H^+_{ab}=p_ap_b-q_aq_b$
and the amplitude $A_+$ at the Newtonian order as
\beq
A_+=\frac{2G^{5/3}\mch_z^{5/3}}{Dc^4}(\pi f)^{2/3}(1+\cos^2i),\label{b5}
\eeq
where $\mch_z$ is the redshifted chirp mass of the binary. For this mode
the principle polarization angle $\psi_{AB}$ is given by
\beq
\tan \psi_{AB}=\frac{(\vex_{A}-\vex_B)\cdot \veq}{(\vex_{A}-\vex_B)\cdot\vep}.\label{b6}
\eeq
The cross ($\times$) mode has the amplitude 
\beq
A_\times=\frac{4G^{5/3}\mch_z^{5/3}}{Dc^4}(\pi f)^{2/3}\cos i,\label{b7}
\eeq
with the polarization tensor $H^\times_{ab}=p_aq_b+q_ap_b$. 
Its  principle polarization angle differs by $\pi/4$ from the plus
mode. 

Now we can write down the phase shift (\ref{b1}) with various parameters of the
binary. For notational simplicities we define a function $U(t,Y)$ that
contains information of the phase of the gravitational wave at the detector $Y$ and time $t$
(taking its origin appropriately) as
\beq
U(t,Y)=\exp[2\pi i f (t+\vex_Y\cdot\ven/c)].
\eeq
Then the quantity $V$ is the real part of the following expression (Eq.(3))
\beqa
& &
\frac12 \lmk\cos2\psi_{AB} A_+ +i\sin 2\psi_{AB} A_\times \rmk \nonumber\\
& \times& \lkk (1-\cos
\theta_{AB}) U(t,A)+2\cos\theta_{AB} U(t-\tau,B)-(1+\cos\theta_{AB}) U(t-2\tau,A) \rkk. 
\eeqa
The expression in the square bracket of the above result can be written as
\beq
u(t,A) R(\theta_{AB},f\tau),
\eeq
where we have defined the factor $R$ as
\beq
R(\theta_{AB},f\tau) \equiv \lkk (1-\cos\theta_{AB})+2\cos\theta_{AB} \exp\lnk2\pi i f \tau
(-\cos\theta_{AB}-1) \rnk -(1+\cos\theta_{AB})\exp(-4\pi i f \tau) \rkk .\label{b11}
\eeq
It is a simple task to  obtain
 the Fourier transformation of the phase shift $\delta \phi_{AB}$. 
In a similar manner  we can make the expression for
 the phase shift of 
 another arm {\it e.g.} $\delta \phi_{AC}$. Then we obtain the signals such as $\delta \phi_{AB}-\delta \phi_{AC}$.

When  gravitational wave-length is much  smaller than the arm-length
 ($f\tau\gg 1$),  the factor
 $R$ shows complicated response that depends strongly on both the angle
 $\ven$ and the frequency $f$. In the long wave approximation ($f\tau\ll 1$) we
 can perturbatively expand Eq.(\ref{b11}), and obtain
\beq
R=4\pi i f \tau \sin^2\theta_{AB}. \label{lla}
\eeq
Now the factor $R$ becomes very simple. 

Parameters such as $\mch_z$, $\mu_z$, $\beta$ are closely related to the
time evolution of the frequency (chirp signal) due to gravitational
radiation reaction as given in Eq.(\ref{pev}). Thus the complicated response of the factor $R$ has
weaker impact on  estimation of these parameters than the direction of the
source $\ven$ (see figure 2). Distance $D$ or the orientation $\vel$ of the
binary  are
determined by using  the
information of the amplitudes $\cos2\psi_{AB} A_+$ and $\sin2\psi_{AB}
A_\times$. 
As seen in the expressions (\ref{b2})(\ref{b4})(\ref{b5})(\ref{b6})(\ref{b7}), they are strongly correlated to the
direction $\ven$ of the binary. Thus their estimation errors are also
improved by the effect of the factor $R$.

%\if0
%%We have used rather  
%As commented in the main text,
% the
%parameter estimation errors $\Delta \gamma_i$ given in table 2 of
%Ref.\cite{Cutler:1998ta} (obtained with using the angular pattern
% function $F^+$ and $F^\times$) agree with  our results by the long wave
% approximation with eq.(\ref{lla}) for the factor $R$.
%\fi

%\input{ref.tex}

\begin{figure}[h]
 \begin{center}
 \epsfxsize=14.cm
 \begin{minipage}{\epsfxsize} \epsffile{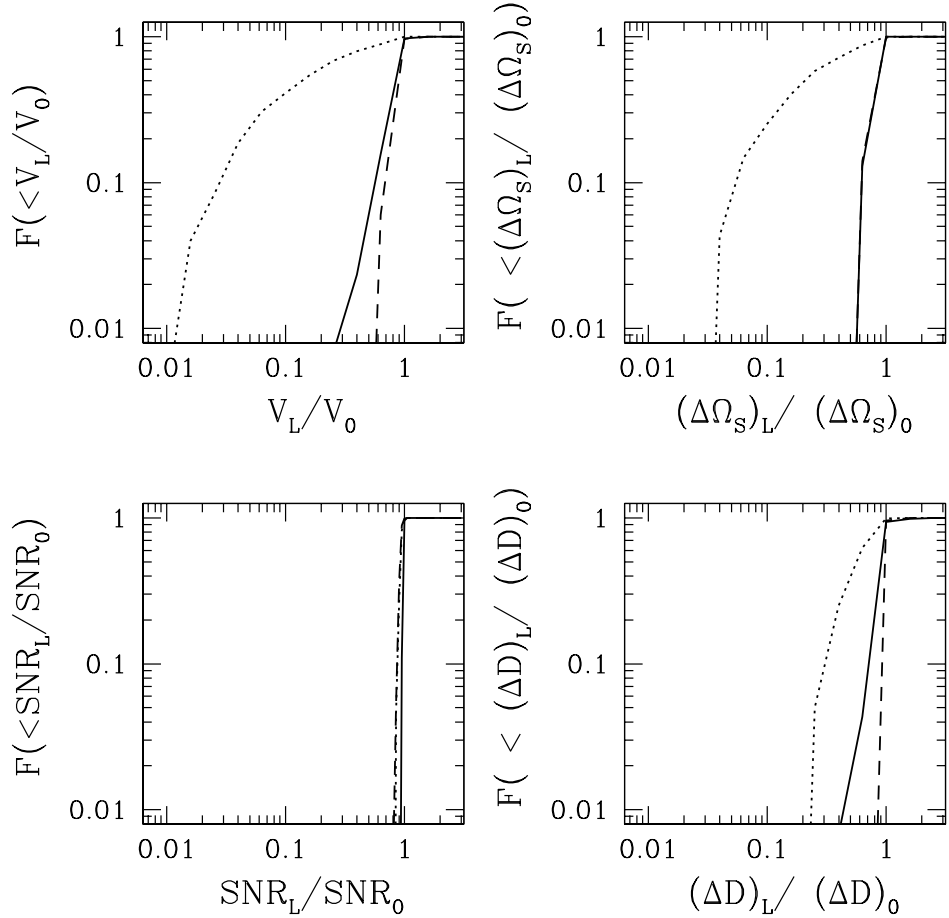} \end{minipage}
 \end{center}
\caption[]{ Distribution of  relative magnitude of the
 three dimensional error box $\Delta V$, $SNR$, the angular resolution
 $\Delta\ven$, and the distance error $\Delta D/D$. We compare
 results from two data streams (with suffix $L$) and those obtained by
 past simple method (with suffix $0$).  The solid lines are
 results for MBH binaries with redshifted masses
 $10^6+10^6 \sol$, the doted lines for $10^5+10^5 \sol$,  and dashed
 lines for  $10^4+10^4 \sol$. We have analyzed 300 binaries
 with random directions and orientations.} 
\end{figure}

\begin{figure}[h]
 \begin{center}
 \epsfxsize=14.cm
 \begin{minipage}{\epsfxsize} \epsffile{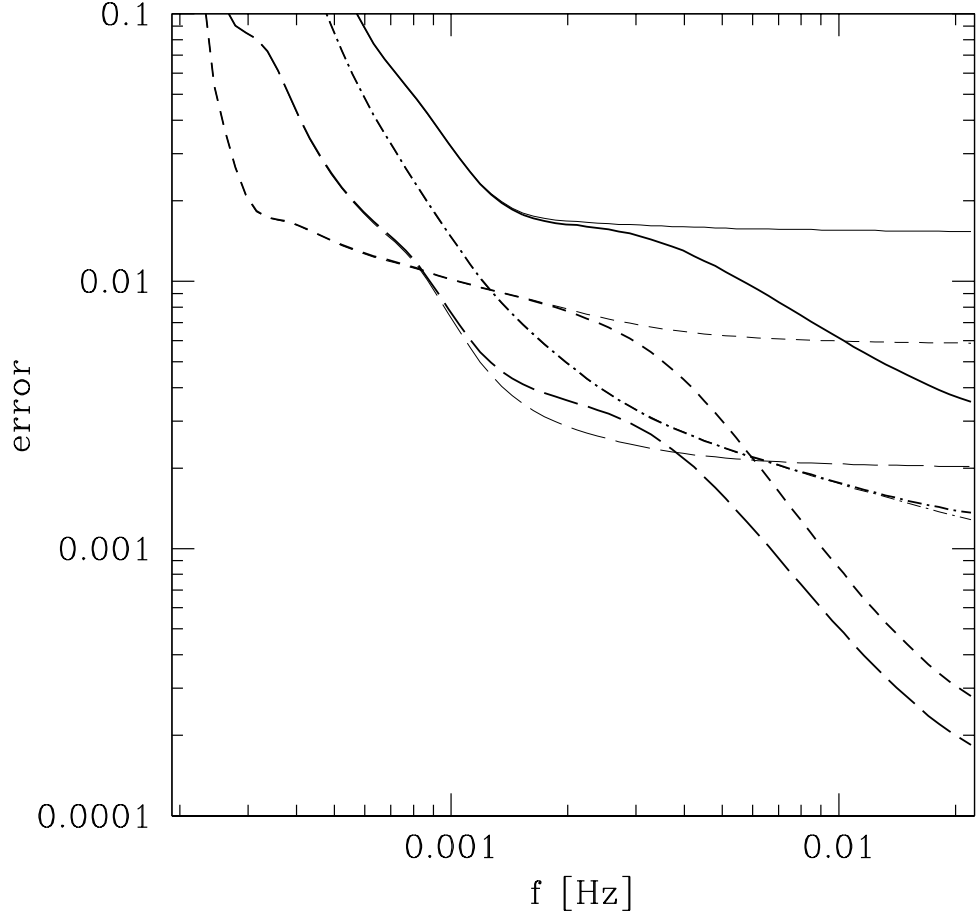} \end{minipage}
 \end{center}
\caption[]{ Dependence of the parameter estimation errors
 $\Delta\gamma_i$ on 
the upper cut-off frequency $f_{cut}\le f_{isco}$. We star integration
 of the signal from $f_0=1.9\times 10^{-4}$Hz. The MBH binary has
 redshifted  
masses $10^5+10^5\sol$, and exists at $z=1$ with
 direction 
 $\theta_s=2.49, \phi_s=0.03$ and orientation $\theta_s=2.32, \phi_s=4.46$. 
The thin lines are the past estimations and thick ones are the new estimations. The solid lines represent for  $\Delta D/D$, the long dashed lines for $\Delta \vel$, the short-dashed lines for $\Delta \ven$, and the dash-dotted lines for the reduced mass $\Delta \mu_z/\mu_z$.  SNR becomes $\sim 1054$ for $f_{cut}=7.0\times 10^{-3}$Hz and $\sim 1140$ for $f_{cut}=2.0\times 10^{-2}$Hz.}
\end{figure}

\begin{figure}[h]
 \begin{center}
 \epsfxsize=14.cm
 \begin{minipage}{\epsfxsize} \epsffile{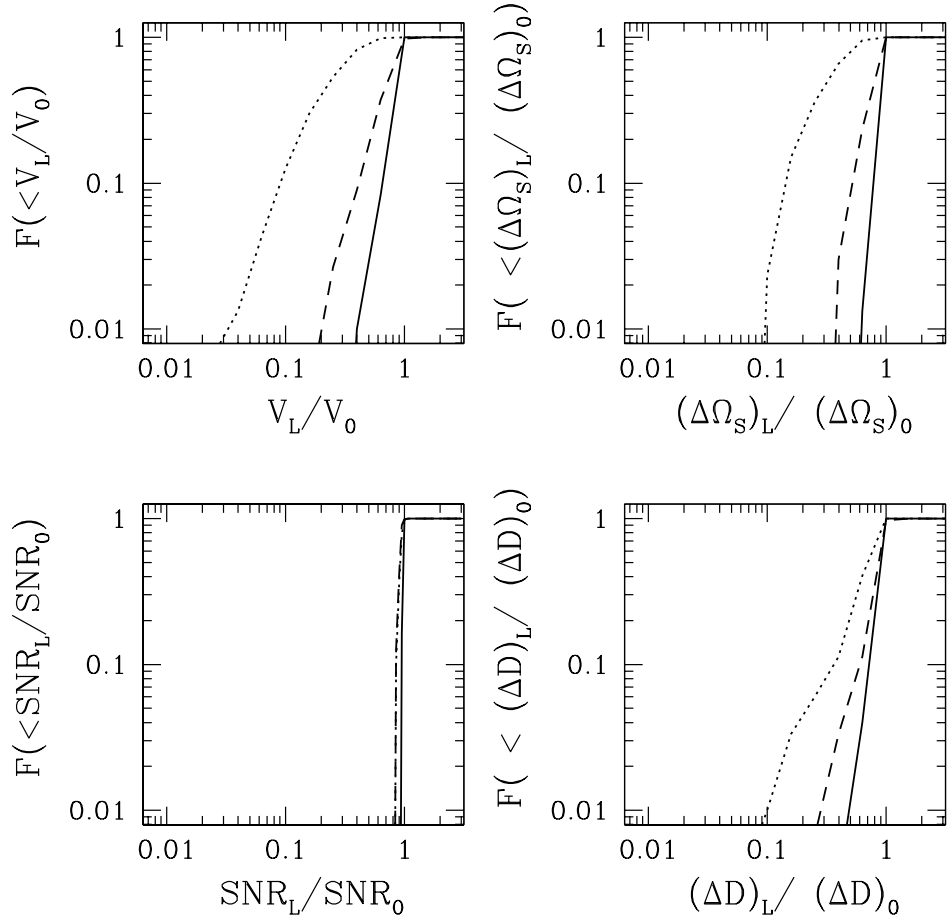} \end{minipage}
 \end{center}
\caption[]{ Same as figure 1, but for the noise canceling combination $X$.} 
\end{figure}

\end{document}